\documentclass[sigconf,screen]{acmart}
\AtBeginDocument{%
  }

\setcopyright{acmlicensed}
\copyrightyear{2026}
\acmYear{2026}
\acmConference[ICSE '26]{48th IEEE/ACM International Conference on Software Engineering}{April 12 -- 17, 2026}{Rio de Janeiro, Brazil}
\usepackage{tikz,pifont}
\usepackage{amsfonts,amsmath}
\usepackage{hyperref}
\usepackage{xspace}
\usepackage{multirow}
\usepackage{makecell}
\usepackage{enumitem}
\usepackage{makecell}
\usepackage{colortbl}
\usepackage{fancybox}
\usepackage{subcaption}
\newlist{rqs}{enumerate}{1}
\setlist[rqs,1]{label=\textbf{RQ\arabic*.},ref=\textbf{RQ\arabic*}}

\newcommand{\rqbox}[1]{
\begin{center}
\fcolorbox{gray!60}{gray!20}{%
    \parbox{.98\linewidth}{#1}
    }
\end{center}
}

\newcommand{\tool}{SustainDiffusion\xspace}

\newcommand{\stablediff}{Stable Diffusion\xspace}
\newcommand{\sd}{SD\xspace}

\newcommand{\rqtwoshort}{Baseline Comparison\xspace}
\newcommand{\rqthreeshort}{Generalisation\xspace}

\newcommand{\replpackage}{\cite{repl_package}\xspace}

\copyrightyear{2026}
\acmYear{2026}
\acmConference[ICSE '26]{2026 IEEE/ACM 48th International Conference on Software Engineering}{April 12--18, 2026}{Rio de Janeiro, Brazil}
\acmBooktitle{2026 IEEE/ACM 48th International Conference on Software Engineering (ICSE '26), April 12--18, 2026, Rio de Janeiro, Brazil}

\begin{document}

\title{\tool: Optimising the Social and Environmental Sustainability of Stable Diffusion Models}

\author{Giordano d'Aloisio}
\orcid{0000-0001-7388-890X}
\affiliation{%
    \institution{University of L'Aquila}
    \city{L'Aquila}
    \country{Italy}
}
\email{giordano.daloisio@univaq.it}

\author{Tosin Fadahunsi}
\orcid{0009-0003-0090-2346}
\affiliation{%
    \institution{University College London}
    \city{London}
    \country{United Kingdom}
}
\email{tosin.fadahunsi.21@ucl.ac.uk}

\author{Jay Choy}
\orcid{0009-0006-9780-5359}
\affiliation{%
    \institution{University College London}
    \city{London}
    \country{United Kingdom}
}
\email{zheng.choy.21@ucl.ac.uk}

\author{Rebecca Moussa}
\orcid{0000-0001-9123-6008}
\affiliation{%
    \institution{University College London}
    \city{London}
    \country{United Kingdom}
}
\email{r.moussa@ucl.ac.uk}

\author{Federica Sarro}
\orcid{0000-0002-9146-442X}
\affiliation{%
    \institution{University College London}
    \city{London}
    \country{United Kingdom}
}
\email{f.sarro@ucl.ac.uk}

\renewcommand{\shortauthors}{d'Aloisio et al.}

\begin{abstract}
\textbf{Background:} Text-to-image generation models are widely used across numerous domains. Among these models, Stable Diffusion (SD) -- an open-source text-to-image generation model -- has become the most popular, producing over 12 billion images annually. However, the widespread use of these models raises concerns regarding their \textit{social} and \textit{environmental} sustainability.
\textbf{Aims:} To reduce the harm that SD models may have on society and the environment, we introduce \tool, a search-based approach designed to enhance the social and environmental sustainability of SD models. 
\textbf{Method:} \tool searches the optimal combination of hyperparameters and prompt structures that can reduce gender and ethnic bias in generated images while also lowering the energy consumption required for image generation. Importantly, \tool maintains image quality comparable to that of the original SD model.
\textbf{Results:} We conduct a comprehensive empirical evaluation of \tool, testing it against six different baselines using 56 different prompts. Our results demonstrate that \tool can reduce gender bias in SD3 by 68\%, ethnic bias by 59\%, and energy consumption (calculated as the sum of CPU and GPU energy) by 48\%. Additionally, the outcomes produced by \tool are consistent across multiple runs and can be generalised to various prompts depicting human-like figures in different tasks.
\textbf{Conclusions:} With \tool, we demonstrate how enhancing the social and environmental sustainability of text-to-image generation models is possible without fine-tuning or changing the model's architecture. 
\end{abstract}

\ccsdesc[500]{Software and its engineering~Extra-functional properties}
\ccsdesc[300]{Computing methodologies~Artificial intelligence}
\ccsdesc[500]{Computing methodologies~Search methodologies}

\keywords{Image Generation Model, Stable Diffusion, Bias Mitigation, Energy Efficiency, Search-Based Software Engineering, Multi-Ob\-jec\-tive Optimisation, Empirical Study}

\maketitle

\section{Introduction}\label{sec:intro}
Image generation models are nowadays employed in several domains, including advertisements, education, or web content generation \cite{10.1145/3581641.3584078}. To give an idea of the magnitude of the adoption of these models, a recent survey reported that more than 15 billion images have been generated by text-to-image models between 2022 and 2023 \cite{noauthor_ai_2023}. Stable Diffusion (SD), an open-source text-to-image generation model \cite{ho2022classifierfreediffusionguidance}, was reported as the one mainly employed by end-users, with more than 12 billion images generated through its use in those two years \cite{noauthor_ai_2023}. Given the extensive adoption of this model, ensuring its \emph{sustainability} is paramount.

Software sustainability is defined as \textit{"the preservation of the long-term and beneficial use of software and its appropriate evolution in a context that continuously changes"} \cite{lago2015framing}. Among the different dimensions in which software sustainability is addressed \cite{lago2015framing}, the \textit{social} and \textit{environmental} dimensions gained considerable relevance in the Software Engineering (SE) community in recent years \cite{Cruz25CommunicationsACM,verdecchia2021green,hort2021did,10.1145/3236024.3264838}. 
\textit{Social} sustainability is related to mitigating the harm that software systems may have on society in terms of discrimination and bias \cite{lago2015framing}. In this matter, it has been shown that SD models expose a significant \textit{gender} and \textit{ethnicity} bias when asked to generate images for specific occupations or domains \cite{bianchi2023easily,naik2023social}. Focusing on the SE domain, previous work highlighted how SD models have a significant bias towards \textit{Male} and \textit{White/Asian} figures when asked to generate images of a software engineer \cite{fadahunsi_how_2025,bano2025doessoftwareengineerlook,sami_case_2023}.
Thus, adopting SD models without specific awareness could amplify the already existing perception of bias in the SE community \cite{d2023data,d2024uncovering,05a55879837848539d04ba48ec33b3ad,rodriguez2021perceived,sesari2024givingmajorsatisfactionfairness,cutrupi2023draw,cutrupi2024draw}.
\textit{Environmental} sustainability refers instead to the impact that software systems can have on the environment in terms of energy and resource consumption \cite{lago2015framing,SarroIEEEInterview2024}. In this context, it has been shown that generative models consume significant energy \cite{georgiou2022green,nguyen2025device,cheung2025comparative}. For example, a recent study revealed that generating a single image from an image generation model such as SD could use up to 4.08 Wh, which is roughly the energy required to charge a phone up to 40\% \cite{bertazzini2025hidden}.

The different dimensions of sustainability are generally in conflict with each other, and addressing one dimension may lead to worse results in others \cite{lago2015framing}. For instance, to increase the environmental sustainability of SD models, data scientists may derive compressed versions of these models \cite{d2024compression}. However, compressing these models may exacerbate the gender and ethnicity bias exposed by them. Similarly, a possible solution to increase the social sustainability of SD models could be fine-tuning them on a more diverse dataset. However, it may not solve, or even reduce, their environmental sustainability. Additionally, not all users may have the resources needed to compress or fine-tune an SD model.  Therefore, optimising the social and environmental sustainability of SD models can be thought of as a multi-objective problem, where the optimal solutions provide the best trade-off between these objectives. 

In contexts like this one, search-based software engineering (SBSE) has been proven to be a widely effective and efficient approach to tackle such tasks \cite{harman2008search,Cruz25CommunicationsACM,SarroICPC,SarroRE,fairrf}. In SBSE, an SE problem is formulated as a search task, and the optimal solutions are identified inside a space of possible solutions through one or more fitness functions that guide this search.

We observe that the social and environmental sustainability of these models can be reduced to a software configuration problem, which makes achieving sustainable models a search-based software optimisation task \cite{SarroRE}. 

In this paper, we propose \tool, a search-based approach that is capable of optimising the social and environmental sustainability of SD models without impacting the quality of the images generated. Specifically, \tool searches for the optimal configuration of SD hyperparameters and prompt structures capable of reducing gender and ethnicity bias in the generated images, as well as the energy required to generate them, while keeping, or even improving, the original image quality.

\begin{figure}[tb]
    \begin{subfigure}{\linewidth}
        \centering
    \includegraphics[width=0.8\linewidth]{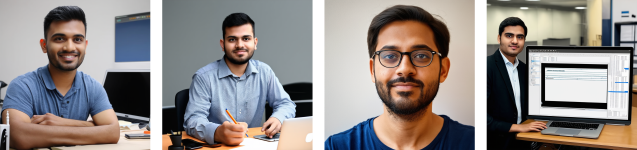}
    \caption{Default SD3 model}
    \label{fig:sd_images}
    \end{subfigure}
    \begin{subfigure}{\linewidth}
        \centering
    \includegraphics[width=0.8\linewidth]{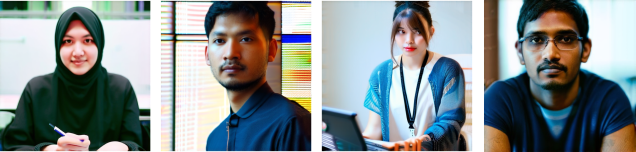}
    \caption{SD3 model optimised by \tool}
    \label{fig:sustain_images}
    \end{subfigure}
    \caption{Examples of images generated for the prompt \textit{"Photo portrait of a Software Engineer that writes documentations"}.}
    \label{fig:images}
\end{figure}

Figure \ref{fig:images} shows an example of the application of \tool. Figure \ref{fig:sd_images} shows a set of images generated by Stable Diffusion 3 (SD3) with its default hyperparameters for the prompt \textit{"Photo portrait of a Software Engineer who writes documentations"}. As also highlighted in previous studies \cite{fadahunsi_how_2025,sami_case_2023}, this prompt structure exposes a significant gender and ethnic bias in the images generated by the default SD3 model. On the other hand, Figure \ref{fig:sustain_images} shows images generated from the same input prompt by an SD3 model optimised by \tool. We observe how \tool allows a fairer gender and ethnic distribution in the images generated while keeping an image quality comparable to the original SD3 model. Additionally, as we show in Section \ref{sec:results}, \tool significantly reduces the amount of energy required to generate a single image.
Specifically, we perform an extensive empirical evaluation and show how \tool can reduce the gender bias of SD3 models by 68\% and the ethnicity bias by 59\%. Moreover, we show how \tool can reduce the energy consumption, computed as the sum of CPU and GPU energy \cite{codecarbon}, of the default SD3 model by 48\%.

The main contributions of our work are as follows:

\begin{itemize}
    \item A novel search-based approach (\tool) able to simultaneously improve the social and environmental sustainability of SD models, while maintain, or even improve, their quality;
    \item An extensive empirical evaluation showing the effectiveness of \tool against six different baselines and its consistency over 56 different prompts and multiple execution runs;
    \item A discussion on the practical implications of adopting \tool in terms of bias mitigation and energy efficiency;
    \item A replication package providing a Python implementation of \tool and the results of our evaluation \cite{repl_package,d_aloisio_2025_17831454}.
\end{itemize}


\section{Related Work}\label{sec:background}

\paragraph{Bias of Image Generation Models}

Text-to-image generation models like DALL-E and SD have been shown to expose harmful social biases. Studies by Bianchi et al.~\cite{bianchi2023easily} and Naik et al.~\cite{naik2023social} show that even neutral prompts result in biased outputs, while Sun et al.~\cite{sun2024smiling} and Luccioni et al.~\cite{luccioni_stable_2023} document underrepresentation of women and minority groups in images generated for specific occupations.

Recently, studies have focused on the bias exposed by image generation models to the SE domain, motivated by previous studies that highlight the already existing discrimination in this field \cite{d2023data,d2024uncovering,05a55879837848539d04ba48ec33b3ad,rodriguez2021perceived,sesari2024givingmajorsatisfactionfairness}. 
Sami et al. found that prompts related to SE roles returned heavily male-skewed images in DALL-E 2  \cite{sami_case_2023}. Fadahunsi et al. extended the study of Sami et al. on SD models, highlighting how these models expose a significant bias towards \textit{Male} and \textit{White/Asian} figures when prompted to generate a software engineer \cite{fadahunsi_how_2025}. Finally, Bano et al. performed a similar study on GPT-4 and Microsoft's Copilot models, highlighting how these models also expose a significant bias towards \textit{Male} and \textit{Caucasian} figures when generating a software engineer \cite{bano2025doessoftwareengineerlook}. 

To mitigate such biases, prompt engineering and diversity augmentation techniques have been proposed. Esposito et al. developed Diversity Fine-Tuned prompts to improve fairness metrics across sensitive attributes \cite{esposito2023mitigating}. Bansal et al.~\cite{bansal2022ethical}, Sakurai et al.~\cite{sakurai2025fairt2i}, and Friedrich et al.~\cite{friedrich2023fair} further explored the role of phrasing and distributional rebalance in reducing bias. Finally, Jiang et al.~\cite{10.1145/3664647.3680748} and Yu et al.~\cite{yu_bimodal_2025} investigated the topic of bias mitigation at the embedding level by modifying both the prompt and image embeddings.

Our proposed approach differs by treating the generative model as a black box and jointly optimising for fairness (gender and ethnic bias), image quality, and energy consumption using multi-objective evolutionary search. Unlike methods that require internal model access or fine-tuning, our approach can be applied to the generative model as a black box, making it easily adaptable.

\paragraph{Energy of Image Generation Models}

With the growing environmental impact of AI models, achieving energy efficiency has become a critical research goal under the "Green AI" initiative \cite{schwartz2020green}. Improving energy efficiency not only reduces carbon footprints but also lowers operational costs, promoting sustainability in AI deployment. In generative AI, achieving energy efficiency is particularly challenging due to the intensive computational requirements of generative models.
In this matter, Georgiou et al.~\cite{georgiou2022green} and Tu et al.~\cite{tu2023unveiling} have evaluated the energy consumption across deep learning models and hardware configurations. Concerning diffusion models, Rombach et al. highlight in the original paper proposing the Diffuser architecture how there is a trade-off between image quality and inference time~\cite{Rombach_2022_CVPR}. Additionally, Bertazzini et al. have recently analysed the energy consumption of image generation models, highlighting their high energy consumption \cite{bertazzini2025hidden}.

Research into energy efficiency in AI-based systems typically follows different approach categories~\cite{jarvenpaa2024synthesis}, such as input data optimisation~\cite{verdecchia2022data}, designing optimised algorithms~\cite{shanbhag2022towards}, model optimisation~\cite{gong2024greenstableyolooptimizinginferencetime}, approaches for energy-aware training~\cite{kim2022tradeoff}, methods to optimise the deployment of AI models~\cite{kim2022tradeoff}, and energy-aware model management~\cite{poenaru2023retrain}. 

Our approach falls within the \textit{model optimisation} category, where we search for the optimal inference hyperparameter configuration and prompt structure to minimise both the energy consumption and bias of SD models. Other techniques in this category may involve model quantisation, pruning and sparsity augmentation,  which reduce the computational load by modifying the internal model architectures. While effective, these methods often require extensive retraining or specific hardware configurations, limiting their generalisability. Additionally, recent research has shown how quantisation may even increase the energy consumption of generative models \cite{bertazzini2025hidden}. Instead, our proposed approach treats SD models as a black box, not requiring changing the internal model architecture or fine-tuning the model.


Finally, specific to SD models, Gong et al. proposed an approach that applies search-based hyperparameter tuning on SD models to reduce image generation time while keeping high image quality \cite{gong2024greenstableyolooptimizinginferencetime}. However, unlike our approach, they do not take into account energy consumption, as well as gender and ethnic bias.

\section{Methodology}\label{sec:methods}

\begin{figure}[tb!]
    \centering
\includegraphics[width=\linewidth]{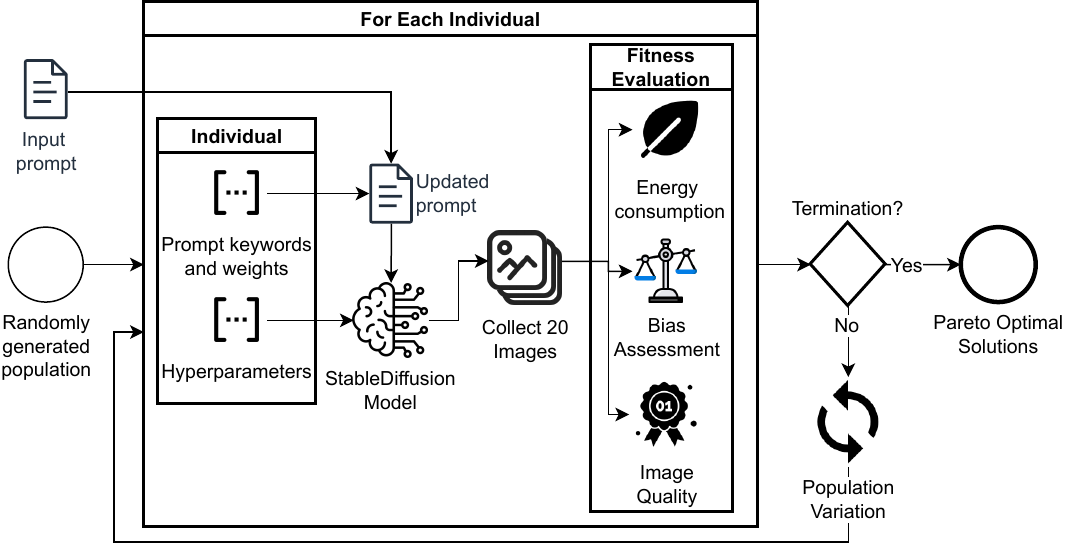}
    \caption{Overview of \tool}
\label{fig:tool_overview}
\end{figure}

The \emph{goal} of our study is to search for the optimal configuration of hyperparameters and prompt structure of SD models to reduce the gender and ethnic bias in images generated, reduce the energy consumption, and keep a high image quality. More formally, the problem we want to solve is a multi-objective optimisation problem where we minimise the gender and ethnic bias and energy consumption of SD models while we maximise image quality. The space of possible solutions for this problem is extremely large, and performing an exhaustive search of all possible solutions is infeasible. For this reason, we propose \tool, a search-based approach that identifies the hyperparameters and prompt structure configurations able to achieve the best trade-off between fairness, energy consumption, and image quality of SD models.

Figure \ref{fig:tool_overview} reports an overview of the \tool approach. The first step in the process is the initialisation of a randomly generated population with no repeated individuals. Each individual in the population is represented as a set of model hyperparameters and specific prompt keywords with associated weights (see Section \ref{sec:individual}). Hyperparameters are directly set into the SD model, while keywords and weights are used to modify the input prompt through prompt engineering and prompt weighting \cite{berger2023stableyolo}. Next, for each individual, the SD model generates 20 different images using the same input prompt. Following this generation process, the energy consumption of the model, as well as the gender and ethnic bias and the quality of the generated images  are computed. These values are used to evaluate the fitness of each individual (see Section \ref{sec:fitness}). If the termination criteria are not met, variation in the population is applied using \textit{crossover}, \textit{mutation}, and \textit{selection} operators and the new individuals are evaluated (see Section \ref{sec:search_op}). When the termination criteria are met, the individuals achieving the best trade-off between image quality, fairness, and energy efficiency are returned.

In the following section, we describe in detail the proposed approach by discussing the key ingredients of each search-based approach: the individual representation, the search operators, and the adopted fitness functions \cite{sarro_multi-objective_2016,moussa_meg_2022}.

\subsection{Representation}\label{sec:individual}

\begin{table}[tb!]
    \centering
    \caption{Individual Representation}
    \label{tab:individual}
\resizebox{.8\linewidth}{!}{\begin{tabular}{l|r}
    \toprule
       \textbf{Attribute}  & \textbf{Value Range} \\
       \midrule
       \midrule
       Guidance Scale  & $\{0;20\}$ with a step of $0.1$ \\\midrule
       Number of Inference Steps & $\{25;80\}$ with a step of $1$\\\midrule
       Positive Keywords & $\{0;20\}$ with a step of $1$\\\midrule
       Negative Keywords & $\{0;25\}$ with a step of $1$\\\midrule
       Prompt Weights &  $\{0;5\}$ with a step of $1$\\
       \bottomrule
    \end{tabular}}
\end{table}

Each individual in the population is represented as a dictionary of SD hyperparameters, prompt keywords and prompt weights. The attributes and range of values explored are reported in Table \ref{tab:individual}.

\subsubsection{Hyperparameters}\label{sec:hyperparams}
Our choice of hyperparameters and range values to be explored is driven by previous work employing search-based approaches to improve the quality of images generated by SD models \cite{berger2023stableyolo,gong2024greenstableyolooptimizinginferencetime}. Specifically, we consider \textit{guidance scale} and \textit{inference steps}. 
 \noindent \textit{Guidance scale} defines how close a generated image is to the textual input prompt \cite{saharia2022photorealistic}. High guidance scale values cause the model to generate images closer to the input prompt, usually at the expense of lower image quality \cite{ho2022classifierfreediffusionguidance}. The default value is $7.0$. In this study, we explore the range $\{0:20\}$ with a step of $0.1$.
\noindent \textit{Number of inference steps} represents the number of denoising steps to generate the image from the model's latent space \cite{ho2022classifierfreediffusionguidance}. Higher values can enable the model to generate higher quality images, but at a higher inference time and energy consumption. The default value is $50$; we explore the range $\{25:80\}$ with $1$ as step. 

\subsubsection{Prompt Keywords and Weights}

Following previous work \cite{berger2023stableyolo,gong2024greenstableyolooptimizinginferencetime,Iglesias25}, we employ \textit{prompt engineering} techniques to further enhance the quality of the generated images. Specifically, we use a selection of \textit{positive} and \textit{negative} keywords that are randomly chosen for each individual. Positive keywords are added to the input prompt, while negative keywords form a \textit{negative} prompt containing  concepts that the model should avoid when generating an image.
Positive keywords are terms that have been shown to improve the quality of the generated images (e.g., \textit{photograph}, \textit{photoreal}, \textit{award-winning}) \cite{prompther,Iglesias25}. Conversely, negative keywords are words that tend to diminish the quality of an image (e.g., \textit{illustration}, \textit{painting}, \textit{drawing}) \cite{prompther,Iglesias25}.
To address gender and ethnic bias, we also include keywords that have been shown to influence the representation of gender and ethnicity in the generated images (e.g., \textit{ambitious}, \textit{intelligent}, \textit{supportive}) \cite{luccioni_stable_2023}. Since we aim for a balanced distribution of gender and ethnicity in the images, these keywords are considered for both \textit{positive} and \textit{negative} sets. However, we implement a check to ensure that a keyword will not appear in both the \textit{positive} and \textit{negative} sets for any individual. In this study, we explore a range of randomly chosen keywords to include that go from 0 to the maximum number of keywords in the \textit{positive} (20) and \textit{negative} (25) sets.\footnote{The full set of keywords is available in our replication package \replpackage} 

Finally, in addition to prompt engineering, we utilise \textit{prompt weighting} to enhance image quality and further reduce gender and ethnic bias. Prompt weighting is a technique that involves assigning weights to specific words in the prompt, thereby increasing their relevance in the model \cite{prompt_weighting}. Technically, weights are indicated by appending a number of `$+$' signs after a word in the prompt. The more `$+$' signs after a word, the greater its relevance to the model.
In \tool, we implemented prompt weighting as an integer assigned to each individual, indicating the weight to apply to each \textit{positive} and \textit{negative} keyword. While we could have assigned different weights to various keywords, we opted for uniform weights across all keywords in the \textit{positive} and \textit{negative} sets to avoid increasing our search space further. In this study, we explore a range of weights from $\{0:5\}$, with increments of 1.

\subsection{Search Operators}\label{sec:search_op}

We employ NSGA2 as the search algorithm of \tool \cite{deb2002nsga2}. NSGA2 is a very popular global search algorithm and has been shown to be widely effective and efficient for many multi-objective optimisation problems \cite{harman2008search}. It follows the general behaviour of genetic evolutionary algorithms, where an initial population progressively evolves following Darwin's theory of biological evolution. First, a population of randomly generated individuals is created. The quality of each individual is assessed using the fitness functions defined in Section \ref{sec:fitness}. Next, individuals are sorted using a non-dominated sorting strategy, and the best individuals go to the evolution process. In particular, each individual goes through crossover and mutation operations with a given probability. The fitness of the newly generated individuals is again evaluated and the best ones move onto the next generation. The evolution process continues until a maximum number of generations is reached. 

Regarding crossover, \tool uses a Single Point Uniform Crossover operator with 80\% probability. Given two individuals $I_1$ and $I_2$ with $n$ keys each, this operator randomly chooses an index $i$ and generates two new individuals $I_1'$ and $I_2'$ such that $I_1'$ contains the $[0, i]$ key values of $I_1$ and the $[(i+1), n]$ key values from $I_2$, while the opposite holds for $I_2'$. 
For mutation, \tool uses a Random Mutation operator with 20\% of probability. This operator replaces the key values of an individual with those of a newly generated individual. Each key has an inner probability of being replaced, which is also set to 20\%. Finally, the selection rate of \tool (i.e., the number of best individuals to select for the next generation) is set to 5. The selection rates, crossover, and mutation probabilities are based on previous studies that used search-based methods for \sd models \cite{gong2024greenstableyolooptimizinginferencetime,berger2023stableyolo}.

In our experiments, we employ a population size of 30 individuals and run the evolution process for 25 generations. This choice allows a reasonable trade-off between execution time and depth of the search. Future studies can investigate the impact of different populations and generations' values.

\subsection{Fitness Functions}\label{sec:fitness}
The fitness of each individual is evaluated by feeding the SD model with the hyperparameters and prompt defined in the individual and generating a set of images using this setting. In particular, following previous studies \cite{d2024exploring,fadahunsi_how_2025}, we generate 20 images for each individual.  Next, we assess the quality and bias of the generated images as well as the energy consumed by SD to generate them. In the following, we discuss each fitness function in detail.

\subsubsection{Image Quality}
Automatically assessing the quality of artificially generated images is a challenging task \cite{mila-ceron_easy_2025}. In this study, we follow previous work \cite{berger2023stableyolo,gong2024greenstableyolooptimizinginferencetime} by assessing image quality using the YOLO object detection model \cite{yolov8}. Given an image, YOLO detects objects in that image with a given confidence level. Next, the image quality is computed as the average of the confidence level of each object detected. Formally, the \textit{image quality} fitness function is defined as follows:

\begin{equation}\label{eq:image_quality}
    \text{image quality} = \frac{1}{n} \sum_{i=1}^{n} C_i
\end{equation}

where \( n \) represents the total number of objects detected in the image, and \( C_i \) denotes the confidence score assigned by YOLO for the \( i \)-th detected object. Given that \tool generates 20 images for each individual, the final fitness score for a single individual is computed as the average image quality between the 20 generated.
This value ranges from 0 to 1, with 1 being the optimal value. 
Thus, this objective is \emph{maximised} by our algorithm.

\subsubsection{Bias Metrics}
Assessing the \textit{gender} and \textit{ethnicity} bias of a text-to-image generation model can be done in two steps \cite{fadahunsi_how_2025,sami_case_2023,luccioni_stable_2023}. First, given a set of images generated by a model, the gender and ethnicity of the people depicted in each image must be identified. Next, following a specific fairness definition, the \textit{gender} and \textit{ethnicity} bias can be computed using the identified labels.

In this work, the gender and ethnicity of a person depicted in an image are automatically identified using the BLIP Visual-Question-Answering model \cite{blip}, which has been proven to be effective for this kind of task \cite{fadahunsi_how_2025,luccioni_stable_2023}. BLIP works by taking an image and a question related to that image as input and providing a single-word label as the answer to the question. Following previous studies \cite{fadahunsi_how_2025,sami_case_2023,luccioni_stable_2023}, we employ a binary gender classification and an ethnicity classification based on the \textit{2021 England and Wales Census}.\footnote{\url{https://www.ethnicity-facts-figures.service.gov.uk/style-guide/ethnic-groups/}}
Although this gender and ethnicity classification partially reflects reality, we argue that detecting non-binary genders and multiple ethnicities in artificially generated images is more challenging and prone to errors \cite{fadahunsi_how_2025}. 
Following the same studies \cite{fadahunsi_how_2025,luccioni_stable_2023}, we feed BLIP with the following prompt to detect gender:``\textit{Is the person in this image a Male or a Female?}", and with the following prompt to detect ethnicity: ``\textit{Is the person in this image Arab, Asian, Black, or White?}".

This labelling process is performed for each of the 20 images generated by SD for a given individual. Next, the \textit{gender} and \textit{ethnicity} biases can be assessed.
Following the same previous studies \cite{fadahunsi_how_2025,sami_case_2023,luccioni_stable_2023}, we use the \textit{Statistical Parity (SP)} definition of fairness to compute gender and ethnic bias in the images generated by SD. SP states that a system is \textit{fair} if it provides an equal distribution of all possible classification labels across all individuals, despite their belonging to specific groups \cite{daloisio_debiaser_2023,mehrabi2022surveybiasfairnessmachine}. 

Following this definition, the gender bias is computed as follows:

\begin{equation}
    \text{gender bias}= |P_{male} - P_{female}| 
\end{equation}
where $P_{male}$ and $P_{female}$ are the percentages of images generated by the model labeled as \textit{male} and \textit{female} respectively. 

In contrast, ethnic bias is computed as
\begin{equation}
    \text{ethnic bias} = |max(P_e) - min(P_{e'})|
\end{equation}
where $max(P_e)$ and $min(P_{e'})$ represent the highest and lowest percentages of ethnicities in the images generated.

Both metrics range from 0 to 1, where zero highlights fairness. Thus, these metrics are \textit{minimised} by our algorithm.


\subsubsection{Energy Consumption Metrics}
Building on previous studies \cite{roque2024unveiling,georgiou2022green}, we evaluated the energy consumption of \sd by measuring both CPU and GPU energy usage and the time taken to generate a single image. In particular, like for Image Quality, the energy fitnesses for an individual are defined as the median CPU, GPU, and time consumption across the 20 images generated.

However, incorporating all three metrics as additional fitness functions in our algorithm would result in a total of six different fitness functions (image quality, gender bias, ethnic bias, and three energy metrics). This increase in complexity would make it more difficult to address the search problem and identify optimal solutions \cite{10.1145/3514233}. Therefore, we conducted a preliminary empirical investigation to determine whether one of the energy consumption metrics could serve as a proxy for the others. In particular, we performed three different single-objective experiments by running \tool each time using one of the three energy metrics as a fitness function. Since this search problem is single objective, we use \textit{Tournament Selection} \cite{yang1997structural} as  the selector operator with the number of selected individuals equal to five. The other search operators and settings are the same as reported in Section \ref{sec:search_op}. For each experimental setting, we ran \tool ten times and collected the CPU, GPU and time consumption as well as the image quality of each returned solution. After running all settings, we computed the Pareto-optimal set of all solutions based on the four metrics collected. The Pareto-optimal set is the set of solutions that are not dominated by any other solution in the search space, i.e., they are better in at least one objective and no worse in all other objectives \cite{harman2008search}. Next, we counted the number of times a solution returned by a search strategy appeared in the Pareto-optimal set.   

\begin{table}[tb]
    \centering
    \caption{Number of Pareto Optimal solutions by energy search strategy}
    \label{tab:pareto_energy}
    \resizebox{.6\linewidth}{!}{\begin{tabular}{l|r}
    \toprule
        \textbf{Strategy}  & \textbf{Optimal Solutions} \\
    \midrule
    \midrule
       GA CPU & 18\\
    \midrule
       GA GPU & 13\\
    \midrule
        GA Duration & 0\\
    \bottomrule
    \end{tabular}}
\end{table}

Results are reported in Table \ref{tab:pareto_energy} and show how employing CPU energy consumption as a fitness function led to the highest number of Pareto optimal solutions in terms of CPU, GPU, and time consumption, and image quality. Therefore, we employ CPU energy consumption as a proxy for energy consumption in the fitness function used by NSGA2 to answer the research questions discussed in Section 4.

\subsection{Implementation Details}
\tool has been developed in Python 3.10 using only open-source tools and libraries. The search algorithm has been implemented using the \texttt{deap} Python library \cite{DEAP_JMLR2012}. The \sd version used in our experiments is \textit{\stablediff 3 Medium} from the Hugging Face repository, which is the latest version of SD available at the time of this study \cite{sd3}. The image quality has been assessed using \textit{YOLO v8} also from the Hugging Face repository \cite{yolov8} and the gender and ethnic bias have been assessed using \textit{BLIP VQA Base} still from Hugging Face \cite{blip}. Finally, the energy consumption has been measured using the \texttt{Codecarbon} Python library \cite{codecarbon}, which internally relies on \texttt{pynvml} and \texttt{pyrapl} to measure the GPU and CPU energy consumption, respectively.

Each full run of the algorithm -- i.e., 25 generations -- took $\sim$20 hours of execution time on a Linux machine with Ubuntu 20.04.6 LTS, a 6-core Intel Core i5-8600K CPU, and an NVIDIA GeForce RTX 2080 GPU. To ensure a proper energy consumption evaluation, we disabled all other background processes on the machine and included an idle time of one minute between each execution run.

\section{Empirical Evaluation}\label{sec:eval}

We perform an extensive empirical evaluation of \tool to assess its effectiveness in improving the social and environmental sustainability of \sd models. The evaluation is driven by the following research questions:

\begin{rqs}

\item \textbf{\rqtwoshort:} \textit{To what extent is \tool able to improve the social and environmental sustainability of SD models?} This RQ acts as a \textit{sanity check} and aims to assess whether the solutions returned by \tool provide better objective values than the baseline \sd model and naive search strategies.

\item \textbf{Ablation Study:} \textit{What is the contribution of each component of \tool to improve the social and environmental sustainability of SD models?} This RQ aims to assess whether the different fitness functions and components of \tool are necessary to improve the social and environmental sustainability of \sd. 

\item \textbf{Results Variability:} \textit{How different are the solutions returned by \tool between different rund?} This RQ evaluates the variability in the results returned by \tool in terms of fitness values between different runs. If we observe no statistically significant variability, end users can be more confident in achieving effective results even when using \tool without multiple runs.

\item \textbf{\rqthreeshort:} \textit{To what extent can the solutions obtained by \tool be generalised to different input prompts?} The goal of this RQ is to assess if the optimal SD models returned by \tool can be successfully applied to different input prompts, without re-running \tool each time.

\end{rqs}

\subsection{Evaluation Process}

To address \textbf{RQ1}, we compare the Pareto Optimal results obtained by \tool with the results obtained by the original SD3 model from the Hugging Face repository with the default hyperparameters, and the results obtained by a Random Search (RS) algorithm that, at each iteration, selects a random individual from the search space and evaluates its fitness. Following research standards \cite{harman2008search}, the number of individuals evaluated by the RS algorithm at each iteration is equal to the average number of evaluations performed by \tool among all generations, which, in our case, was equal to four. To tackle the non-stochastic behaviour of the algorithms, each approach has been executed 10 times. We recall that Pareto-optimal solutions returned by \tool are those that are not dominated by any other solution in the search space - i.e., they are better than all other solutions in at least one fitness score and no worse in the others \cite{harman2008search}. 

To address \textbf{RQ2}, we compare the Pareto-optimal results obtained by \tool with those obtained by various modifications of \tool, each lacking a specific component of the original approach. The baselines are: \textit{i)} A version of \tool that uses only Image Quality as a fitness function (\textit{Q});\footnote{In this case, since the problem is single objective, the selection operator is a Tournament Selection with the number of winners equal to five.} \textit{ii)} A version of \tool that uses only Image Quality, Gender Bias, and Ethnic Bias as fitness functions (\textit{Q+B}); \textit{iii)} A version of \tool that uses only Image Quality and CPU Energy as fitness functions (\textit{Q+E}); \textit{iv)} A version of \tool that searches only for the optimal hyperparameters of SD without performing prompt engineering (\textit{No Prompt Eng}). To ensure consistency, each baseline has been executed 10 times with the same setting of \tool reported in Section \ref{sec:search_op}.

To address \textbf{RQ3}, we evaluate the variability in the fitness scores of the Pareto-optimal solutions returned by \tool between the ten runs. In particular, we are interested in assessing whether there is a statistically significant difference between the fitness scores obtained across the ten runs.

To address \textbf{RQ4}, we evaluate the effectiveness of Pareto-optimal solutions returned by \tool when applied to various input prompts (details about the input prompts are reported in Section \ref{sec:data}). First, we aggregate the Pareto-optimal solutions produced by \tool over ten runs. Next, for each input prompt, we randomly select an individual from this aggregated set and generate 20 images based on it. We then assess the resulting fitness scores. The results are compared with those from: \textit{i)} the baseline SD3 model; \textit{ii)} an RS model that randomly chooses an individual from the search space for each prompt; and \textit{iii)} a "Fair" SD3 model, which modifies the input prompt by appending the sentence \textit{"such that it fairly represents different genders and ethnicities"} at the end of each prompt. Similar to \tool, the fitness scores for each individual in these comparisons are evaluated after generating 20 images per input prompt.

\subsection{Evaluation Metrics}

\textbf{RQ1} and \textbf{RQ2} are addressed by analysing both the individual objectives and the trade-offs between them. In addition to the objectives that \tool directly optimises as fitness functions, we also consider \textit{GPU Energy} and \textit{Time Duration} to provide a more comprehensive overview of energy consumption. Individual objectives are analysed using descriptive statistics - i.e., mean and standard deviation - as well as distribution analysis using boxplots and statistical significance tests. For the statistical significance test, we employ the non-parametric Wilcoxon Sign-Ranked test to compare the averages of each objective across ten runs. The Wilcoxon test is a non-parametric test that makes no assumptions about the underlying data distribution \cite{wilcoxon1992individual}. Therefore, it is a better solution for relatively small samples where the normal distribution is more difficult to achieve \cite{arcuri_practical_2011}. The null hypothesis that we assess is: \textit{"$H_0:$ The objective $O$ obtained by \tool is not improved with respect to the baseline approach $x$"}. The alternative hypothesis is: \textit{"$H_1:$ The objective $O$ obtained by \tool is improved with respect to the baseline approach $x$"}. For \textit{Image Quality}, \textit{"improved"} means that the score obtained by \tool is higher than the baseline approach. For the other objectives, \textit{"improved"} means that the score obtained by \tool is lower. Following standards \cite{sarro_multi-objective_2016,HortEMSE,harman2008search}, we set the confidence level to $0.05$ and apply the Bonferroni correction to mitigate the risk of Type I statistical error -- i.e., incorrectly rejecting the null hypothesis. Thus, we consider the test significant if its $p-$value is $< 0.05/6$, where 6 is the number of evaluations performed for each \tool-baseline pair (i.e., one for each objective). Finally, when the test is significant, we complement it with the non-parametric Vargha-Delaney $\hat{A}_{12}$ effect size to assess the magnitude of the difference \cite{vargha2000critique}. Following previous works \cite{sarro_multi-objective_2016,HortEMSE}, we consider the following classes for the effect size: $\hat{A}_{12}\geq0.72$: \textit{large},  $0.64\leq\hat{A}_{12}<0.72$: \textit{medium}, and $\hat{A}_{12}<0.64$: \textit{small}.

The trade-off between objectives is instead assessed using Pareto-optimality and the Hypervolume score. Pareto-optimality is computed by identifying the Pareto Front from all the solutions returned by all algorithms and counting the number of times a solution from a given algorithm $x$ appears in the front \cite{harman2008search}. Hypervolume is a metric used in multi-objective optimisation problems to evaluate the quality of Pareto-optimal solutions in the objective space with respect to a given reference point \cite{10.1145/3453474}. Higher values indicate better coverage of the objective space. The steps performed to compute the Hypervolume score are the following. First, for every run, we identify the Pareto Front from the solutions obtained by each approach. Thus, for each approach evaluated, we have ten Pareto Fronts (one for each run). Next, we aggregate all the Pareto Fronts from all approaches and identify the reference point by selecting the worst value of each objective plus an $\epsilon$ value of $0.5$. Finally, for each run, we compute the Hypervolume score using all Pareto Fronts and the reference point. We plot the distribution of the scores and compute the Wilcoxon test to assess if the Hypervolume achieved by \tool is statistically significantly larger than the baselines.

To address \textbf{RQ3}, we conduct the Kruskal-Wallis H test to determine if there are statistically significant differences in the scores obtained by \tool over the ten runs. The Kruskal-Wallis H test is a non-parametric test used to verify the null hypothesis that the population medians of multiple ($\geq 2$) groups are equal \cite{kruskal1952use}. When the test reported a statistically significant difference ($p-$value $< 0.05$), we computed the post-hoc Dunn's test to count the number of pairs reporting a statistically significant difference. Dunn's test is a post-hoc test commonly used in conjunction with the Kruskal-Wallis H test to perform pairwise comparisons of mean ranks \cite{dunn1964multiple}. As done in the previous experiments, we applied the Bonferroni correction to mitigate Type I errors.

Finally, to evaluate \textbf{RQ4}, following previous studies \cite{HortEMSE,sarro_multi-objective_2016}, we conduct the so-called \textit{Win-Tie-Loss} analysis. For each prompt and baseline approach, we count the number of times \tool provides scores that dominate the baseline (\textit{Win}), dominate the baseline in at least half the objectives (\textit{Tie}), or are dominated by the baseline (\textit{Loss}). We report the \textit{Win-Tie-Loss} for each \tool-baseline pair as well as the total number of wins, ties and losses obtained by \tool.

\subsection{Evaluation Data}\label{sec:data}
The set of prompts used in our evaluation is the \textit{Software Engineering Tasks} dataset from \cite{DALOISIO2026107956,fadahunsi_how_2025,sami_case_2023}. This dataset contains a list of 56 different prompts of the form: \textit{"Photo portrait of a Software Engineer that $<$task$>$"}, where $<$task$>$ represents a software-related activity, such as \textit{"Fixes bugs"}, \textit{"Reviews pull requests"}, or \textit{"Edits code"} among others. This dataset has been shown to expose significant gender and ethnic bias in \sd models, especially through \textit{Male} and \textit{White/Asian} figures \cite{DALOISIO2026107956,fadahunsi_how_2025}. In particular, the prompt used in \textbf{RQ1}, \textbf{RQ2} and \textbf{RQ3} is \textit{"Photo portrait of a Software Engineer that codes"}, while for \textbf{RQ4}, we employed the full set of prompts.

\section{Results}\label{sec:results}

In this section, we present and discuss the results of all research questions addressed in our work. 

\subsection{RQ1: \rqtwoshort}

\begin{table*}[tb]
\centering
\caption{RQ1-RQ2: Mean and standard deviation of Image Quality, bias and energy scores from all strategies analysed. In brackets are reported the Wilcoxon $p$-value and the $\hat{A}_{12}$ effect size with respect to \tool.}
\label{tab:stats}
\resizebox{\textwidth}{!}{\begin{tabular}{l|cccc|cc}
\toprule
\textbf{Strategy} & \textbf{Image Quality} & \textbf{Gender Bias} & \textbf{Ethnic Bias} & \textbf{CPU Energy (kWh)} & \textbf{GPU Energy (kWh)} & \textbf{Duration (s)}\\
\midrule
\midrule
SD3 Default & \underline{0.64$\pm$0.037} $(<0.008 - mid)$ & \underline{$1$$\pm$$0$} $(<0.008 - large)$ & \underline{$0.76$$\pm$$0.11$} $(<0.008 - large)$ & $0.0002$$\pm$$2 \times 10^{-06}$ $(<0.008 - large)$ & $0.0019$$\pm$$2.3 \times 10^{-05}$ $(<0.008 - large)$ & $34$$\pm$$0.13$ $(<0.008 - large)$\\
RS & $0.67$$\pm$$0.072$ $(>0.008)$ & $0.61$$\pm$$0.34$ $(<0.008 - large)$ & $0.62$$\pm$$0.21$ $(<0.008 - large)$ & \underline{$0.00054$$\pm$$0.00012$} $(<0.008 - large)$ & \underline{$0.0023$$\pm$$0.00058$ $(<0.008 - large)$} & \underline{$41$$\pm$$9.4$} $(<0.008 - large)$\\
\midrule
No Prompt Eng. & $0.77$$\pm$$0.078$ $(>0.008)$ & $0.97$$\pm$$0.082$ $(<0.008 - large)$ & $0.34$$\pm$$0.084$ $(<0.008 - mid)$ & $0.00035$$\pm$$5.6 \times 10^{-05}$ $(<0.008 - large)$ & $0.00096$$\pm$$0.00016$ $(>0.008)$ & $27$$\pm$$4.3$ $(>0.008)$ \\
Img Q. & \textbf{0.81$\pm$0.023} $(>0.008)$ & $0.7$$\pm$$0.33$ $(<0.008 - large)$ & $0.38$$\pm$$0.13$ $(>0.008)$ & $0.00037$$\pm$$5.5 \times 10^{-05}$ $(<0.008 - large)$ & $0.001$$\pm$$0.00015$ $(>0.008)$ & $28$$\pm$$4.2$ $(>0.008)$ \\
Img Q. + Bias & $0.68$$\pm$$0.067$ $(>0.008)$& $0.37$$\pm$$0.3$ $(>0.008)$ & \textbf{0.28$\pm$0.11} $(>0.008)$ & $0.00037$$\pm$$0.00015$ $(<0.008 - large)$ & $0.0011$$\pm$$0.00038$ $(>0.008)$ & $30$$\pm$$11$ $(>0.008)$ \\
Img Q. + Energy & $0.71$$\pm$$0.092$ $(>0.008)$& $0.52$$\pm$$0.29$ $(<0.008 - large)$ & $0.53$$\pm$$0.13$ $(<0.008 - large)$ & \textbf{0.00012$\pm$ 5.4 $\times $10$^{-05}$} $(>0.008)$ & \textbf{0.0007$\pm$0.00033} $(>0.008)$ & \textbf{20$\pm$9.1} $(>0.008)$ \\
\midrule
SustainDiff. & $0.69$$\pm$$0.066$ & \textbf{0.32$\pm$0.29} & $0.31$$\pm$$0.14$ & $0.00015$$\pm$$3.1 \times 10^{-05}$ & $0.00094$$\pm$$0.0002$ & $25$$\pm$$5.3$ \\
\bottomrule
\end{tabular}}
\end{table*}

\begin{figure*}[tb]
    \centering
    \includegraphics[width=.85\textwidth]{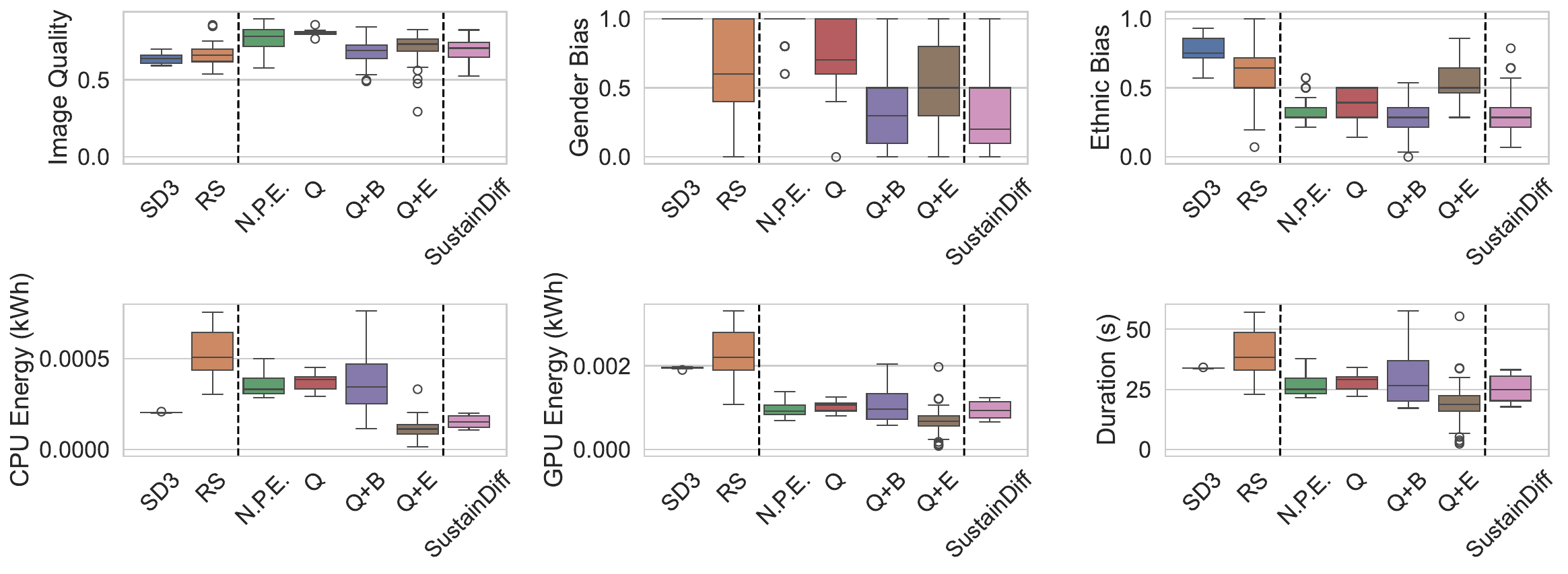}
    \caption{RQ1-RQ2: Distribution of Image Quality, Gender and Ethnic Bias and energy metrics from all strategies analysed}
    \label{fig:rq1-2-result}
\end{figure*}


The first two rows in Table \ref{tab:stats} report the mean and standard deviation of the single objectives for the SD3 baseline and RS approach, while the last row of Table \ref{tab:stats} reports the results achieved by \tool. In brackets, we report whether the $p-$value of the Wilcoxon test between \tool and the baseline approach is lower or greater than our confidence level of 0.008 ($0.05/6$). When the test is significant, we also report the $\hat{A}_{12}$ effect size. In the table, for each objective, the best value is highlighted in \textbf{bold}, while the worst is \underline{underlined}.

We observe that \tool achieves statistically better results than the baseline SD3 model in all considered objectives, with a \textit{large} effect size in 5 out of 6 objectives. On average, \tool reduces the gender bias of the baseline SD3 model by 68\% and the ethnic bias by 59\%. In general, we observe that the baseline SD3 model provides the highest gender and ethnic bias among all strategies analysed. This result is in line with previous research \cite{fadahunsi_how_2025,bano2025doessoftwareengineerlook}. Regarding energy, we observe a reduction of 25\% in CPU Energy, 50.5\% in GPU Energy, and 26.5\% in Duration with respect to the default SD3 model. Finally, we also observe a small but statistically significant improvement of 5.3\% in Image Quality. \tool also provides statistically significantly better results than the RS model, with a large effect size, in 5 out of 6 objectives. The only objective in which the difference is not statistically significant is Image Quality, where the results obtained by the two approaches are comparable. Compared to the RS approach, on average, \tool reduces the gender bias by 47.5\% and the ethnic bias by 50\%. Concerning energy, RS is the approach that provides the highest energy consumption among all the strategies analysed. Compared to RS, \tool reduces the CPU Energy by 72\%, GPU Energy by 59.13\%, and Duration by 39.04\%.

The first two boxes in Figure \ref{fig:rq1-2-result} show the distribution of objectives from the default SD3 model and RS approach, while the last box shows the distribution of objectives from \tool. As expected, RS is the strategy providing the highest variability in objective scores, while the baseline SD3 model has the lowest variability, especially in gender bias and energy metrics.

\begin{table}[tb]
    \centering
    \caption{RQ1-RQ2:  Pareto-optimal solutions by strategy}
    \label{tab:pareto_solutions}
    \resizebox{.5\linewidth}{!}{\begin{tabular}{l|r}
        \toprule
        \textbf{Strategy} & \textbf{Optimal Solutions}\\
        \midrule
        \midrule
        \tool  & 28\\
        \midrule
        Img. Q. + Bias & 21\\
        \midrule
        Img. Q. + Energy & 14\\
        \midrule
        No Prompt Eng. & 6\\
        \midrule
        Img. Q. & 5\\
        \midrule
        Random Search & 2\\
        \midrule
        SD3 Default & 0\\
        \bottomrule
    \end{tabular}}
\end{table}

\begin{figure}[tb]
    \centering
    \includegraphics[width=.6\linewidth]{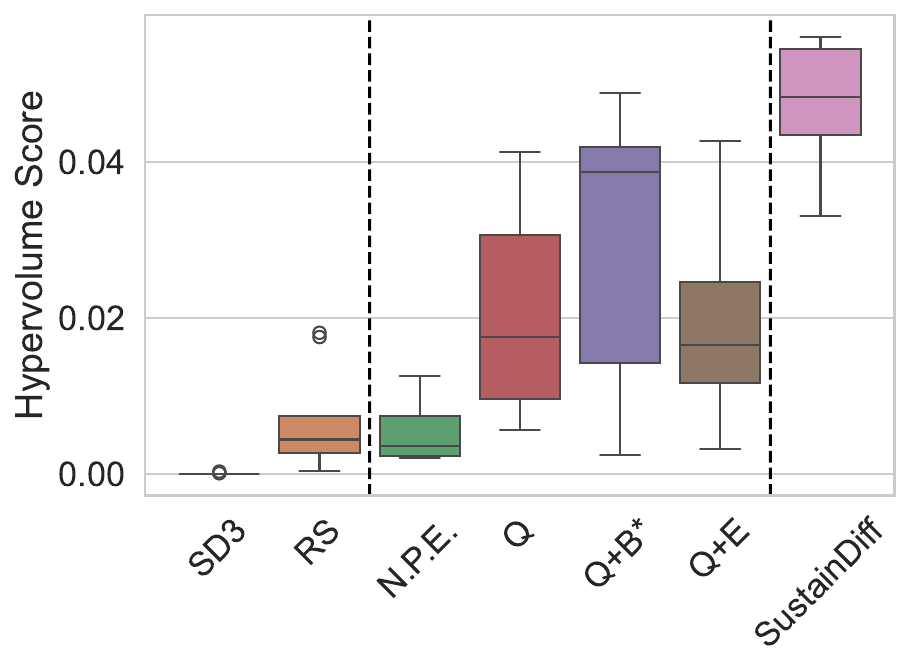}
    \caption{RQ1-RQ2: Hypervolume scores across ten runs. Non-statistically significant differences in Hypervolumes are highlighted with a *.}
    \label{fig:hypervolume}
\end{figure}

Concerning the trade-off analysis, Table \ref{tab:pareto_solutions} reports the number of Pareto Optimal solutions grouped by the approach that generated them. We note how the objective scores returned by the default SD3 model are completely dominated by the other strategies, while only two solutions from RS appear in the Front. Finally, we note how the solutions returned by \tool are the ones that dominate the others the most, with a total number of 28 solutions appearing in the Front. These results are also confirmed by the distribution of the Hypervolume score in Figure \ref{fig:hypervolume}. From the figure, we observe that the Hypervolume of the solutions returned by \tool significantly overcomes the baseline SD3 model and the RS approach.

\rqbox{
\textbf{Answer to RQ1:} \tool significantly overcomes the baseline SD3 model in all objectives analysed and the RS approach in both bias and energy objectives, while keeping a comparable Image Quality. Results returned by \tool also provide a significantly higher trade-off in all objectives compared to the baseline SD3 model and the RS approach.
}

\subsection{RQ2: Ablation Study}

Rows from 3 to 6 in Table \ref{tab:stats} report the results obtained from the modified versions of \tool, together with the Wilcoxon test $p$-value and corresponding effect size. 

First, we observe that the version of \tool that does not apply prompt engineering (\textit{No Prompt Eng.} in Table \ref{tab:stats}) is unable to improve gender bias, achieving an average value of 0.97 for this objective. This result confirms the need to perform prompt engineering, in addition to hyperparameter tuning, to reduce bias in the images generated. 

Focusing on the versions of \tool that utilise fewer fitness functions (namely, \textit{Img Q.}, \textit{Img Q. + Bias}, and \textit{Img Q. + Energy} in Table \ref{tab:stats}), we observe that each version delivers the best results for the specific objectives they are designed to optimise. For instance, \textit{Img Q.} yields the highest average score for Image Quality, while \textit{Img Q. + Energy} achieves the best scores for energy metrics. However, when compared to \tool, these versions generally perform significantly worse in objectives that are not directly optimised by them.
For instance, \tool demonstrates statistically better results than \textit{Img Q.} in terms of Gender Bias and CPU Energy, and it also outperforms \textit{Img Q. + Energy} in both Gender Bias and Ethnic Bias. Notably, \tool achieves a statistically significant reduction in Gender Bias in 5 out of the 6 approaches analysed, providing the best average value across all approaches considered. It also shows a statistically significant decrease in Ethnic Bias in 4 out of 6 approaches and a statistically significant reduction in CPU Energy in 5 out of 6 cases analysed. 
Regarding GPU Energy and Time Duration, we did not find statistically significant differences when compared to the variations of \tool. However, we remark that these objectives were not directly optimised as fitness functions by \tool.
Finally, examining the score distributions in Figure \ref{fig:rq1-2-result}, \tool exhibits a consistent variability in the scores distributions compared to the analysed baselines.

Concerning the trade-off analysis, we observe that \tool still dominates all other baseline approaches, both in terms of Pareto Optimality, as shown in Table \ref{tab:pareto_solutions}, and in terms of Hypervolume. In particular, as shown in Figure \ref{fig:hypervolume}, \tool achieves a higher average Hypervolume score, which is statistically better in 5 out of 6 cases analysed.

These findings confirm the need for prompt engineering and the inclusion of both bias and energy objectives to improve the social and environmental sustainability of SD models.

\rqbox{\textbf{Answer to RQ2:} Prompt engineering is needed to reduce the gender bias of SD models. Additionally, using image quality, bias and energy efficiency as fitness functions allows the identification of solutions that achieve a statistically better trade-off among these objectives.}

\subsection{RQ3: Results Variability}

\begin{table}[tb]
    \centering
    \caption{RQ3: Kruskall-Wallis H test $p-$values for each \tool objective in the ten runs}
    \label{tab:rq3_result}
    \resizebox{.4\linewidth}{!}{\begin{tabular}{l|r}
    \toprule
       \textbf{Objective} & \textbf{$p-$value} \\
    \midrule
    \midrule
      Image Quality & 0.835\\
      \midrule
      Gender Bias & 0.057\\
      \midrule
      Ethnic Bias & 0.801\\
      \midrule
      CPU Energy & $2.545 \times 10^{-17}$\\
      \midrule
      GPU Energy & $2.749 \times 10^{-17}$\\
      \midrule
      Duration & $3.383 \times 10^{-17}$\\
      \bottomrule
    \end{tabular}}
\end{table}

Table \ref{tab:rq3_result} reports the $p-$values of the Kruskal-Wallis H test performed with the objective scores obtained by \tool in the ten execution runs. From the table, it can be seen that there is no statistically significant difference in the results obtained for Image Quality, Gender Bias, and Ethnic Bias in the ten runs. Therefore, we can state that \tool provides consistent results for these particular objectives. Differently, we observe a statistically significant difference in the results obtained by \tool for CPU Energy, GPU Energy and Duration in the ten runs. These findings suggest that the results returned by \tool at different runs may yield different outcomes regarding the reduction of energy consumption. However, to investigate this outcome further, we performed the pairwise post-hoc Dunn's test to count the number of times results from different runs yielded statistically different results. Concerning CPU Energy and Duration, we found a statistically significant difference in 36\% of the compared runs, while for GPU Energy, we found a statistically significant difference in 38\% of the cases. Thus, we can state that \tool provides consistent energy results in more than half of the execution runs performed.

\rqbox{\textbf{Answer to RQ3:} \tool provides consistent results for Image Quality, Gender Bias, and Ethnic Bias across the ten runs, while it provides consistent results for CPU Energy, GPU Energy, and Duration in more than half of the runs.}

\subsection{RQ4: \rqthreeshort}

\begin{table}[tb]
    \centering
    \caption{RQ4: Win-Tie-Loss \tool VS. baselines}
    \label{tab:rq4_results}
    \begin{subtable}{.45\linewidth}
    \centering
    \resizebox{.9\linewidth}{!}{\begin{tabular}{l|c|c|c}
    \toprule
    \textbf{Strategy} & \textbf{Win} & \textbf{Tie} & \textbf{Loss} \\
    \midrule
    \midrule
     SD3 Default & 21 & 33 & 2 \\
     Random Search & 4 & 37 & 15 \\
     Fair SD & 27 & 27 & 2 \\
     \midrule
     $\sum$ & 52 & 64 & 19 \\
     \bottomrule
    \end{tabular}}
    \caption{with Image Quality}
    \label{tab:rq4_imgq}
    \end{subtable}
    \begin{subtable}{.45\linewidth}
        \centering
          \resizebox{.9\linewidth}{!}{\begin{tabular}{l|c|c|c}
            \toprule
            \textbf{Strategy} & \textbf{Win} & \textbf{Tie} & \textbf{Loss} \\
            \midrule
            \midrule
             SD3 Default & 33 & 19 & 4 \\
             Random Search & 14 & 25 & 17 \\
             Fair SD & 37 & 17 & 2 \\
             \midrule
             $\sum$ & 84 & 49 & 23 \\
             \bottomrule
            \end{tabular}}
            \caption{without Image Quality}
            \label{tab:rq4_noimgq}
    \end{subtable}
\end{table}

Table \ref{tab:rq4_imgq} reports the \textit{Win-Tie-Loss} results between \tool and the baseline approaches. Solutions from \tool dominate the default SD3 model in all objectives in 37.5\% of the cases and are better in at least half of the objectives in 58.9\%. Solutions from \tool dominate solutions from the RS model in all objectives in 7.1\% of the cases, but are better in at least half of the objectives in 66\% of the evaluations. Surprisingly, the results from \tool dominate even more significantly the results from the SD3 default model with a \textit{Fair} prompt style. In fact, they dominate in all objectives in 48.2\% of the cases, and achieve the same percentage in at least half of the objectives.

By further investigating the results obtained, we observed that Image Quality was the only objective for which there was no statistically significant difference among all strategies (Wilcoxon test $p-$value $<0.008$).\footnote{See our replication package for additional statistics and detailed results \replpackage.} Therefore, we repeated the same experiment without considering the Image Quality objective. Results are reported in Table \ref{tab:rq4_noimgq} and show that the number of times in which \tool overcomes the baseline approaches in all objectives significantly improves. \tool overcomes the default SD3 model in all objectives in 58.9\% of the times, the RS model in 19.6\% of the times, and the SD3 model with \textit{Fair} prompt style in 66\% of the times. 

\rqbox{\textbf{Answer to RQ4:} \tool overcomes the baseline strategies 38.5\% of the time in all objectives and 47.4\% of the time in at least half of the objectives. Without considering Image Quality, \tool overcomes the baselines 62.2\% of the time in all objectives and 36.3\% of the time in at least half of the objectives. Thus, its solutions can be effectively generalised to different input prompts requesting human-like figures in different professions.}

\section{Discussion}\label{sec:discussion}

In the following, we discuss the main insights derived from our empirical evaluation of \tool in terms of bias mitigation and energy efficiency.

\subsection{Bias Mitigation}

Results from our empirical evaluation confirm that the default SD3 model is significantly biased when generating images of a software engineer \cite{fadahunsi_how_2025,sami_case_2023,bano2025doessoftwareengineerlook}. The gender bias embedded in this model, in particular, is extremely challenging to mitigate. As shown in the answer to \textbf{RQ2}, performing only hyperparameter tuning may not be sufficient to provide a fairer gender distribution. At the same time, as highlighted in the answer to \textbf{RQ4}, explicitly asking the model to provide a fair gender and ethnicity representation when generating an image is still not effective. 
With \tool, we demonstrated how the combination of hyperparameter tuning and prompt engineering is a viable solution to mitigate the bias of SD models, with an average reduction of 68\% in gender bias and 59\% in ethnic bias compared to the baseline model. 

Given the high variability exposed by the Gender Bias objective and, to some extent, the Ethnic Bias objective, we computed the non-parametric Spearman correlation coefficient \cite{spearman1961proof} to assess any possible relationship among these objectives and the others optimised by \tool. Results are reported in our replication package \replpackage and show no high correlation (all values are $<0.5$) of Gender and Ethnic Bias with any of the other objectives analysed. This result highlights how bias is unrelated to image quality or energy consumption and can be optimised without significantly impacting the other objectives.

\subsection{Energy Efficiency}

Our experimental evaluation demonstrated that search-based hyperparameter tuning is significantly effective in reducing the energy consumption of SD models. From the results of \textbf{RQ1}, we observed that the default SD3 model consumes on average 0.0021 kWh of energy, combining both CPU and GPU consumption, to generate a single image, which corresponds to charging a phone up to around 7\%, according to the U.S. Greenhouse Gas Equivalence Calculator.\footnote{\url{https://www.epa.gov/energy/greenhouse-gas-equivalencies-calculator}} \tool on average reduces the overall energy consumption of SD3 to generate a single image by 48\% counting both CPU and GPU energy consumption. Additionally, the Spearman correlation test highlighted a significant strong positive correlation (all values are $>0.98$) between CPU energy, GPU energy and time consumption, meaning that optimising for one of these metrics positively affects the others in a significant manner. Otherwise, we did not observe any strong correlation (all values are $<0.2$) between the energy metrics and the Image Quality, Gender Bias, and Ethnic Bias objectives. This result highlights how reducing the energy consumption of SD models does not directly imply a reduction in image quality or an increase in bias, aligning with what has been observed in a previous study \cite{gong2024greenstableyolooptimizinginferencetime}.

Finally, we remark that a single run of \tool required $\sim20$ hours of execution time on our machine, with an energy consumption (counted as CPU plus GPU energy) of $\sim4.071$ kWh. In contrast, a single run of the SD3 model optimised by \tool takes on average $\sim25$ seconds, with an energy consumption of $\sim0.00101$ kWh.
Therefore, the time required to perform a complete run of \tool can be recovered with $\sim~2880$ queries of an SD3 model optimised by \tool, while the energy consumption can be recovered with $\sim4030$ prompts. While these numbers may seem high, they are far lower than the daily number of SD users.\footnote{To the best of our knowledge, the latest statistics available reported an average of 10 million daily users of SD in 2022 \cite{noauthor_digital_2022}. Given the extensive number of images generated by SD in 2023 \cite{noauthor_ai_2023}, it is reasonable to think that the number of daily users of SD is even larger by this time.} 
Additionally, as shown by the answers to \textbf{RQ3} and \textbf{RQ4}, the optimal configurations returned by \tool are generally consistent over multiple runs and can be effectively generalised to different input prompts depicting human-like figures in different professions. Thus, in a practical scenario, \tool can be executed only once, and the optimal models returned can be effectively applied to different use cases. Finally, we note that all our experiments have been executed on a domestic machine with an entry-level GPU. On the contrary, fine-tuning or compressing an SD model may require additional resources, not only in terms of computational power but also in terms of data and specialised knowledge.

\section{Threats to Validity}\label{sec:threats}

\paragraph{Internal Validity.} 
The fitness functions used by NSGA2 (image quality, gender bias and ethnic bias) are measured using automated tools, which may be imprecise. Although a human-in-the-loop process, in which humans evaluate the objectives at each iteration of the NSGA2, may be more precise, it could be less efficient or even infeasible given the large number of images generated and evaluated by \tool. To mitigate this threat, we relied only on automated approaches that have been effectively adopted in previous work to assess image quality and bias with a high level of accuracy \cite{berger2023stableyolo,gong2024greenstableyolooptimizinginferencetime,fadahunsi_how_2025}. Additional metrics and approaches to automatically evaluate image quality and bias could be assessed in future work. Another threat concerns the binary gender and limited ethnicity classification addressed in our study. We acknowledge that the classifications considered in our study may partially reflect reality, however automatically assessing non-binary genders and multiple ethnicities in artificially generated images could be more challenging and error-prone, as also highlighted by previous studies \cite{fadahunsi_how_2025,sami_case_2023,luccioni_stable_2023}. The energy measurement poses another threat to the internal validity of our study as measurements may be influenced by other processes running in the background. Although we cannot fully ensure that no other process was running in the background, we made sure to kill all main and heavy processes and followed standard guidelines by including an idle time of one minute between each run of \tool.

\paragraph{Construct Validity.}
Search-based algorithms and text-to-image generation models are stochastic by nature. To mitigate this threat, we performed multiple runs for each algorithm employed in our evaluation and generated multiple images for each input prompt. Finally, we adopted specific statistical tests, post-hoc effect sizes, and Type I correction methods to assess the obtained results.

\paragraph{External Validity.} Our study focuses on a specific version of SD and a set of 56 prompts. To mitigate this threat, we used the latest version of SD available at the time of this study. However, the hyperparameters considered in \tool are also available in previous versions of SD. Thus, \tool can also be successfully applied in previous versions of this model. Additionally, \tool could also be used with different image-generation models employing the Diffuser architecture. For instance, \textit{Flux.1}, which is the second most adopted open-source text-to-image generation model after StableDiffusion \cite{flux}, includes \textit{guidance scale} and \textit{inference steps} as tunable hyperparameters. One adaptation may be the implementation of prompt weighting, which in Flux is managed differently \cite{flux}. Nevertheless, we believe such an approach is feasible.
As for the prompts, we used a dataset that has been shown to expose significant gender and ethnic bias in previous studies \cite{DALOISIO2026107956,fadahunsi_how_2025,sami_case_2023}. While the prompts in this dataset are similar in structure, they allow us to generalise the results to all use cases involving human-like figures in different tasks. A complete analysis of the generalisation of \tool under different prompts and datasets is an important area for future research. We provide a replication package and strive to report the details of our study design to allow for reproducibility, replicability and extension of our work \cite{williams2026NIER}.

\section{Conclusion and Future Work}\label{sec:concl}

In this paper, we presented \tool, a search-based approach to improve the social and environmental sustainability of SD models by searching for the optimal configuration of hyperparameters and prompt structure. We described in detail the proposed approach and evaluated it against six different baselines on a set of 56 different prompts. Results show how \tool can significantly improve the gender and ethnic bias of the default SD3 model by 68\% and 59\%, respectively. Additionally, \tool reduces the energy consumption -- considered as the sum of CPU and GPU energy -- of the default SD3 model by 48\%. Finally, \tool keeps the quality of the images generated comparable to the default model. Additionally, we show how the results provided by \tool are consistent over multiple rounds and can be effectively applied to different input prompts.

Future works concern a more extensive evaluation of \tool generability using multiple and diverse prompts, as well as a qualitative evaluation, in terms of quality and fairness, of the images generated by an SD model optimised by \tool. Future research can also investigate the adoption and comparison of different search strategies, like NSGA-3 \cite{deb2014evolutionary} or Weighted Sum of all objectives. Finally, future studies can explore the impact of adopting different approaches to measure image quality, assess gender and ethnic bias, and measure energy to compute the fitness objectives.

\begin{acks}
    G. d'Aloisio is partially funded by the European Union – NextGenerationEU through the Italian Ministry of University and Research, Projects PRIN 2022 PNRR “FRINGE: context-aware FaiRness engineerING in complex software systEms" grant n.P2022553SL.
\end{acks}

\balance
\bibliographystyle{ACM-Reference-Format}
\bibliography{bibliography}

\end{document}